# The monetary growth order

Günter von Kiedrowski[$] and Eörs Szathmáry[§]

[$] Prof. Dr. Günter von Kiedrowski, Chair of Organic Chemistry I, NC 2/171, Ruhr University of Bochum, Universitätsstr. 150, 44780 Bochum, Germany, Email: kiedro@rub.de

[§] Prof. Dr. Eörs Szathmáry, Parmenides Foundation, Kirchplatz 1, 82049 Munich/Pullach, Germany, Email: Szathmary.Eors@gmail.com

**Abstract:**

Growth of monetary assets and debts is commonly described by the formula of compound interest which for the case of continuous compounding is the exponential growth law. Its differential form is $dc/dt = i\,c$ where $dc/dt$ describes the rate of monetary growth, $i$ the compounded interest rate and $c$ the actual principal. Exponential growth of this type is fixed to be neither resource-limited nor self-limiting which is in contrast to real economic growth (such as the GDP) which may have exponential, but also subexponential, linear, saturation, and even decline phases. As a result assets and debts commonly outgrow their economic fundament giving rise to the financial equivalent of Malthusian catastrophes after a certain interval of time. We here introduce an alternative for exponential compounding and propose to replace $dc/dt = i\,c$ by $dc/dt = i\,c^p$ where the exponent $p$ (called reaction order in chemistry) is a quantity which will be termed *monetary growth order*. For $p = 0$, growth is linear as in a simple, non-compounded interest case. For $p = \frac{1}{2}$ growth is parabolic, for $p = 1$ exponential, and for $p = 2$ hyperbolic. While for the case of exponential growth ($p = 1$) the relative (e.g. percentual) increase of assets and debts only depends on interest rate and time, for all other cases ($p <> 1$) relative increase also depends on the principal. If the monetary growth order is supraexponential ($p > 1$), large assets and debts will outperform smaller ones, both, on an absolute and relative scale. For subexponential growth ($p < 1$), small assets and debts grow faster than large counterparts on a relative scale, while not outperforming large assets and debts absolutely. The monetary growth order $p$ is seen as a tuning handle which enables to adjust gross monetary growth to real economic growth. It is suggested that the central banks take a serious look to this control instrument which allows tuning in crisis situations and immediate return to the exponential norm if needed.

**Introduction**

The differential equation:

(1) $$\frac{dc}{dt} = i\, c^p$$

has the solution

(2) $$c_t = \left[c_0^{1-p} + (1-p)\, i\, t\right]^{\frac{1}{1-p}}$$

where $c_t$ is the capital or principal after time t, $c_0$ the principal at time t = 0, p the monetary growth order and i the interest rate for continuous compounding.

Both equations, in which p generally regulates the "explosiveness" of growth of a quantity c, have been studied and discussed outside economics. They are linked to chemical and biological research dealing with self-replicating systems and the origin of life on earth. Equation (1) in which i is a parameter describing the efficiency of growth was briefly discussed by Eigen and Schuster in their Hypercycle theory whose focus is the integration of genetic information to a systemic whole in which competing genetic replicators have to cooperate in order to achieve a higher level task [1]. Such higher-level task was seen in the fixation of the genetic code leading to the last universal common ancestor. Eigen and Schuster distinguished between a monomial range of growth (0 < p < 1), exponential growth (p = 1) and hyperbolic growth (1 < p <= 2). They arrived at the conclusion that while Darwinian evolution labours on the monomial and exponential cases, the origin and fixation of the genetic code must have been based on an event with hyperbolic characteristics. This work triggered the quest for artificial self-replicating systems in chemistry. In such systems, self-replication is seen as the combination of two principles: autocatalysis and information transfer. Autocatalysis in chemistry is equivalent to compounding in finance – the product of a reaction is also the catalyst of the reaction: more products also mean more catalysis. If a reaction can produce a series of competing autocatalysts, information transfer means that a given autocatalyst ensures that its offspring is of the same kind as itself. This is similar to correct accounting in finance where the interest is compounded to the principal of the same account (as opposed to someone else's account). A first example of a chemical self-replicating system was discovered in 1986: Autocatalysis could be detected in the synthesis of a short piece of DNA, 6 bases long, from two even shorter pieces, each 3 bases long [2]. The kinetic analysis of the system revealed a type of growth which was hitherto unknown in chemistry. The underlying rate law was coined *square root law of autocatalysis*, which simply means equation (1) with a "reaction order" of p = ½ for the concentration c (viz. quantity per volume) of the autocatalytic product. The square root law was confirmed by studies of independent systems [3]. For p = ½ and $c_0 \rightarrow 0$, one expects from equation (2) that the autocatalytic growth is parabolic, which could be demonstrated by an experimental system in 1991 [4]. Note however that while equations (1) and (2) describe resource unlimited growth, any real growth is resource limited which was taken into account by the consideration of mass-balance. A theoretical analysis of such systems revealed that the square root law and parabolic growth result from product inhibition, viz. a part of the product feeds back into its own synthesis while another part is aggregated and thus inactive as a replicator [5]. The relative distribution between inactive and active products depends on the concentration and is strictly determined by the thermodynamic properties of products.

**The effect of the growth law on who survives**

Szathmáry studied the implications of the new type of growth while reconsidering earlier work of Eigen and Schuster [6, 7]. Each growth order p has distinct evolutionary consequences. Consider the general case where two or more replicators compete for the uptake of resources/food under flow equilibrium conditions. Technically, such conditions can be found in a flow reactor with an inflow of "food molecules" and an outflow of product molecules. The reactor itself is a compartment in which self-replicating product molecules consume such food molecules and as a result increase their population size (viz. their concentration). The food is completely converted into replicator molecules and the rate of growth depends on the efficiency i of individual replicators. For the sake of simplicity let us assume that there are just two replicators A and B with efficiency parameters $i_A$ and $i_B$. Their initial concentrations – equivalent to principals – are $c_{0,A}$ and $c_{0,B}$. The question to be addressed: What happens if one starts the inflow of food and waits for a sufficiently long time – as much time as needed to establish stationary equilibrium. The outcome of competition strictly depends on the growth order p:

Let us first consider the case p = 0, viz. no autocatalytic feedback (viz. no compounding in terms of finance) at all. After very long time (t → ∞) the stationary concentrations of two replicators A and B, $c_A$ and $c_B$ are:

(3) $\qquad$ p = 0: $\qquad \dfrac{c_{\infty,A}}{c_{\infty,B}} = \dfrac{i_A}{i_B}$

Note that this outcome does not depend on the initial concentrations – it is just the rate of making new molecules which count on the long run. In chemistry, $c_A/c_B$ relates to selectivity, while $i_A/i_B$ relates to reactivity. For a non-autocatalytic situation which is the usual case in chemistry, selectivity is thus controlled by reactivity. Both competitors will coexist forever populating the reactor in accordance to their efficiency.

Let us now consider a square root law (p = ½ in equation (1)). Independent of the initial concentrations, the outcome is:

(4) $\qquad$ p = ½: $\qquad \dfrac{c_{\infty,A}}{c_{\infty,B}} = \left(\dfrac{i_A}{i_B}\right)^2$

This corresponds to a case of coexistence with increased selectivity. If the efficiency ratio were 1/10 the stationary population ratio would be 1/100. If one understands efficiency as a metrics for fitness, the more efficient and thus stronger replicator disproportionally benefits from the competition, while not pushing away its competitor completely. This outcome is in contrast to the conclusion derived earlier.

For exponential growth (p = 1) the outcome of competition is dramatically different:

(5) $\qquad$ p = 1: $\qquad \dfrac{c_{\infty,A}}{c_{\infty,B}} = 0 \; if \; i_A < i_B$

$\qquad\qquad\qquad\qquad \dfrac{c_{\infty,A}}{c_{\infty,B}} = \infty \; if \; i_B < i_A$

$\qquad\qquad\qquad\qquad \dfrac{c_{\infty,A}}{c_{\infty,B}} = 1 \; if \; i_B = i_A$

This case marks selection in the sense of an Darwinian "*survival of the fittest*". Unless the fitness/efficiency of both replicators is not exactly the same, one replicator is sentenced to go extinct. Stable coexistence is thus not possible. Note that in this setup the outcome again does not depend on the initial concentration of replicators. If the weaker replicator had a higher concentration in the beginning, the establishment of flow equilibrium and the time needed to reach selection is just a bit longer compared to the case where weak started at low concentrations.

Unlimited hyperbolic growth (p = 2) asymptotically approaches infinity at finite times and thus exhibits a singularity. The outcome of competition differs from the previous cases with respect to its dependency from the initial concentrations of competitors:

(6) $\quad$ p = 2: $\quad \frac{c_{\infty,A}}{c_{\infty,B}} = 0 \; if \; i_A c_{0,A} < i_B c_{0,A}$

$$\frac{c_{\infty,A}}{c_{\infty,B}} = \infty \; if \; i_A c_{0,A} > i_B c_{0,A}$$

$$\frac{c_{\infty,A}}{c_{\infty,B}} = 1 \; if \; i_A c_{0,A} = i_B c_{0,A}$$

What counts here is the product of efficiency and initial replicator concentration. If both replicators exhibit identical efficiencies the one with a higher initial concentration will win. In chemistry, the case of identical efficiency is expected for reactions that lead to enantiomers, viz. molecules that differ in their threedimensionality like the left or right hand, or, like an irregular object and its mirror-imaged pendant. In evolution, hyperbolic growth has been implied in "once-and-forever selection" such as the origin of the genetic code [1], due to fixation by "*survival of the most common*".

Already in 1938, in the "Golden Age" of theoretical ecology, Vito Volterra proposed that for sexually reproducing populations the basic growth rate cannot be exponential (Maltusian), since males and females have to find each other. If we disregard sex differences, this implies p = 2, with interesting implication for ecology [8]. Moreover, a group of economists, including Brian Arthur, have analyzed positive feedback in economy. According to his generalization, resource-based firms operate with diminishing and knowledge-based firms with increasing returns [9]. Increasing returns in economy results in the "survival of the common" type of survival, just as in biology.

Concerns about the limitation of natural resources and thus the limits of growth from an economic, social and political perspective were addressed by the Club of Rome [9+]. These thoughts in a way reconsidered and sublimed the concept of non-matched und apparently unbalanced growth leading to Malthusian catastrophes in general. The report was widely recognized as its publication coincided with the oil crisis and the public awareness of oil as a limited natural resource.

**Money as a replicator**

While at the first moment it looks bizarre that results from the study of molecular replication could have implications for finance, a more general look indeed reveals that money may be seen as a replicator. The current finance system is based on the division of labour between a central bank and a set of merchant (commercial) banks. The central bank produces or removes liquidity by printing or destroying bank notes at a rate enabling the stability of prices. It also fixes the prime rate which commercial banks pay as interest to the central bank when they need liquidity. Commercial banks appearantly deal as mediators of exchange between borrowers and lenders. Borrowers pay interest to the bank as cost of borrowing while lenders earn interest from the bank as income from lending. The exchange process itself however is not constrained by a law of conservation: Fiat money is loaned into existence by the process of credit creation paramount to the banking sector. While the credit itself is payed back leading to its erasure, the interest payed by the borrower will ever remain causing the net growth of fiat money. In addition, fractional reserve banking implies that only a fraction of an existing deposit need to be reserved for cash withdrawals – the rest can be loaned out as new credits. Income to the bank thus not only results from the interest rate difference between lending and borrowing as well as for fees and tuition to be paid for the banking service, but also from the fact that the process of credit creation may be and usually is perpetuated. This is more easily understood when considering a customer who deposits his loan from bank A into an account at bank B. Bank B is entitled to see the newly deposited money as a resource for creating new credits not considering that the deposited money was itself the subject of credit creation at another bank before. Lately at bank B, interest appearently arises ex nihilo.

As borrowing yields debts while lending yields assets, the sum of debts is in balance with the sum of assets including those created by the bank for its own capitalization. In principle, assets and debts may grow without limits so long as the bank assures that its own capitalization does not fall below a critical limit fixed by financial regulations. From the simplistic point of view of a chemist and biologist, money is a replicator because assets and debts exhibit autocatalysis by interest compounding while information transfer is fulfilled by the service of proper accounting.

László Mérő has advocated the view that capital is replicator, and it "builds" companies as survival machines (or vehicles) analogous to the way in which genes build organisms. It is not difficult to realize that this is insufficient, since organisms are built under the influence of many different genes cooperating on the same playground: the organism. He proposes that there are specific memes (replicating cultural items) that specifically act together with capital: the mones (alluding to money, memes and genes with this term). The mone is a piece of information describing a trait of a company that can attract capital investment. Many mones cooperate to specify a company (an organism in economy). The business plan of a company is a comprehensive analysis of mones [10].

We are not proposing to demolish knowledge-based economy, the mones of which allow for increasing returns and hence survival of the common. This relates to the survival of companies. We are proposing that interests in the financial sector should follow subexponential growth. This might (is likely?) to dampen the effects of increasing returns, thus knowledge (in an interesting twist) gets re-established in our economy: intrinsic fitness i cannot completely be masked by supraexponential growth: intelligent start-ups would have more chance to invade the market.

As the demand for money is expected to scale with macroeconomic growth data (presumably in a linear relationship) the growth of money is expected to match the "class of explosiveness" observed in such data. So far, central bank steered to reach this match by fixing the prime rate, making the borrowing/lending of money via merchant banks either less or more expensive/profitable. What was out of control by central banks so far was the development of the distribution/spread of assets and debts leading to the polarization between rich and poor in a developed economy. Distributive justice so far is a matter of politics and taxation policies of governments and states, and not of a central bank. Could it be otherwise?

Presumably yes. It is highly likely that the question of coexistence versus selection is generally influenced by the explosiveness of growth, regardless of the nature of the competitors. Biological species competing for a niche, self-replicating molecules competing for survival in a reactor, or assets competing for the best growth rate, are all subjects for Darwinian scenarios enabling evolution in various realms. Human civilizations usually benefit from cooperation within setups which allow for the emergence of multiple winners. When it comes to money, adjustable subexponential growth may be the most acceptable recipe to solve the increasing imbalance between rich and poor. All what is needed is the fine tuning of a single numerical parameter p, the monetary growth order.

**The monetary growth order p**

The exponent p in equation (1) and (2) is easy to comprehend as a parameter which controls the effectiveness of a principal c. For p = 1 small and large principals are equally effective with respect to the relative increase by interest. Exponents p > 1 make large principals more effective than small ones on a relative scale while the opposite is true for exponents p < 1. This is demonstrated by table 1 which shows how increasing principals respond to the cases p = 1.1, p = 1, and p = 0.9.

**Table 1:** The effect of the monetary growth order p on the principal c.

| Principal c | Effective Principal $c^p$ | | |
|---|---|---|---|
| | at p = 1.1 | at p = 1.0 | at p = 0.9 |
| 1,000.00 € | 1,995.26 € | 1,000.00 € | 501.19 € |
| 10,000.00 € | 25,118.86 € | 10,000.00 € | 3,981.07 € |
| 100,000.00 € | 316,227.77 € | 100,000.00 € | 31,622.78 € |
| 1,000,000.00 € | 3,981,071.71 € | 1,000,000.00 € | 251,188.64 € |
| 10,000,000.00 € | 50,118,723.36 € | 10,000,000.00 € | 1,995,262.31 € |
| 100,000,000.00 € | 630,957,344.48 € | 100,000,000.00 € | 15,848,931.92 € |
| 1,000,000,000.00 € | 7,943,282,347.24 € | 1,000,000,000.00 € | 125,892,541.18 € |
| 10,000,000,000.00 € | 100,000,000,000.00 € | 10,000,000,000.00 € | 1,000,000,000.00 € |
| 100,000,000,000.00 € | 1,258,925,411,794.17 € | 100,000,000,000.00 € | 7,943,282,347.24 € |
| 1,000,000,000,000.00 € | 15,848,931,924,611.20 € | 1,000,000,000,000.00 € | 63,095,734,448.02 € |

The only compounding algorithm we know is the one described by equation (1) and (2) for the exponential case (p = 1). There are centuries of experience with this case, may be because it is beautifully simple and apparently transparent to everyone, may be because it is due to the opposite. In the following the exponential case of compounding (p = 1) is compared with supraexponential (Fig. 1a-d), and subexponential cases (Fig. 2a-d). All growth curves shown are calculated using continuous compounding. Data are shown for a period of time which exceeds the historical average of a stable economy. If the growth is exponential and the interest rate is 5% it takes 100 years to increase a given initial capital by a factor of 148. Asset doubling for an interest rate of 100% would take ln(2) = 0.7 years, 7 years at 10%, and 70 years at 1% – the time scaling linearly with the interest rate. At 5% interest rate we expect about 14 years needed for a doubling, and as 148 is little larger than 128 = $2^7$ but smaller than 256 = $2^8$, we estimate the time needed to reach a factor of 148 as little more than $7 \times 14 = 98$ years. Indeed, the scale on the abscissa is as expected. Next, for an exponential compounding we expect that the given factor of 148 should not vary with the principal. This is indeed the case, as can be seen by comparison of Fig. 1a-d, which accounts for a starting capital of 1 €, 1000 €, 1 Mio €, and 1 Mrd. € respectively.

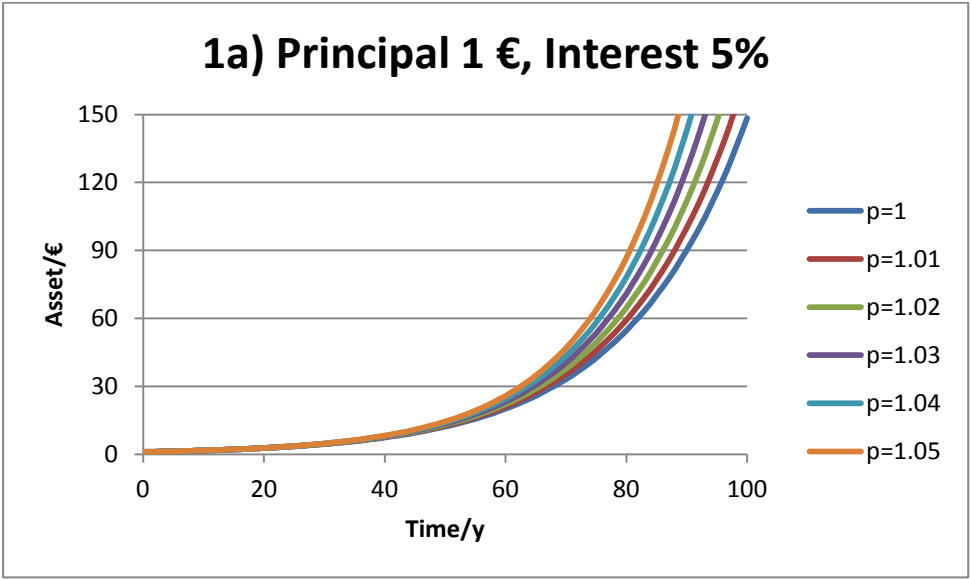

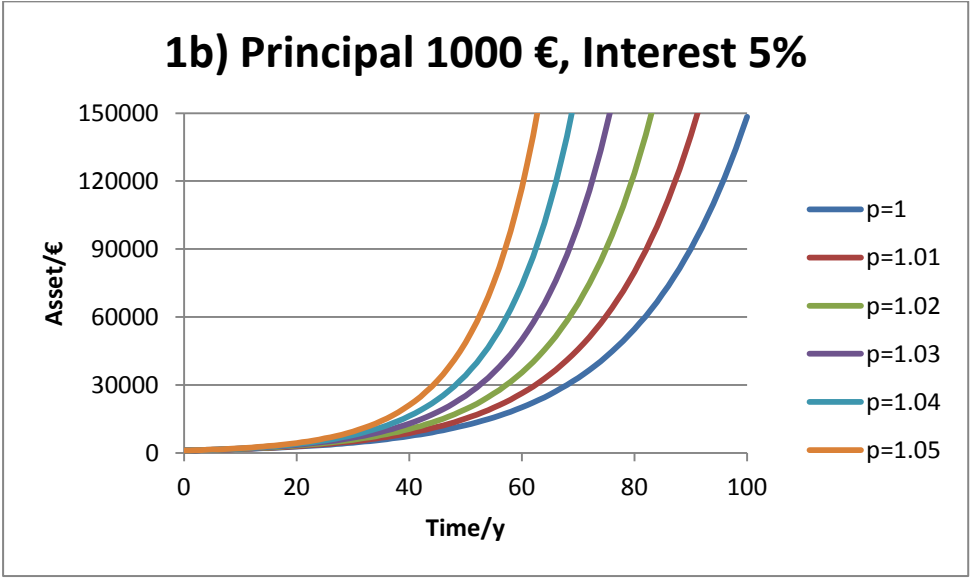

**Figure 1:** Asset growth in the supraexponential case for a principal of (a) 1 €, (b) 1000 €, (c) 1 Mio €, (d) 1 Mrd €, a fixed annual interest rate of 5%, and continuous compounding. Monetary growth orders are shown in the legends.

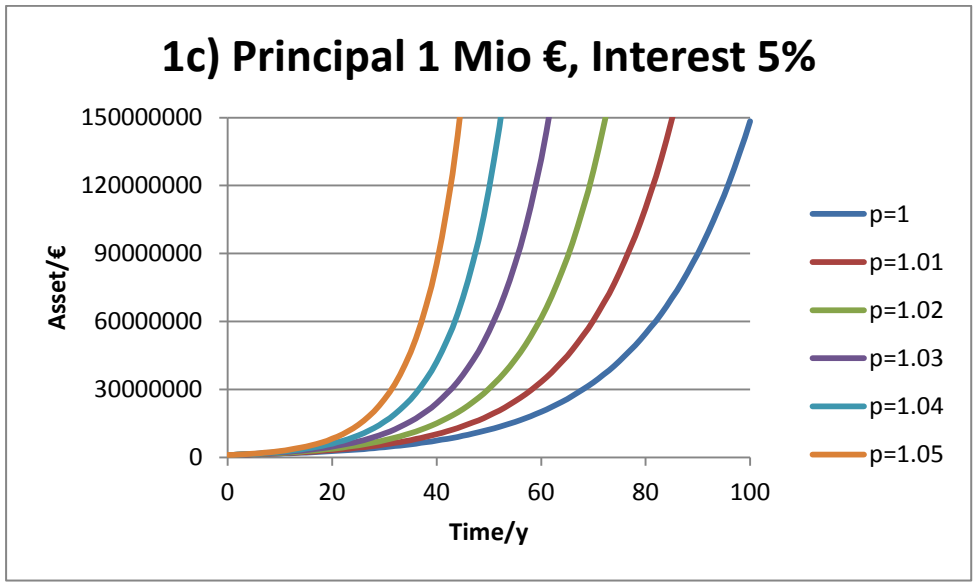

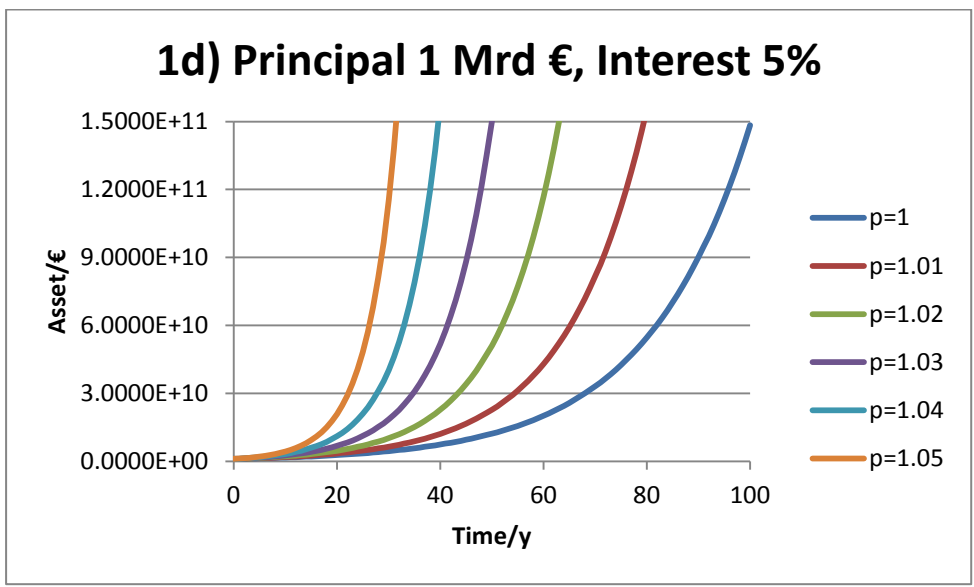

**Figure 1** continued

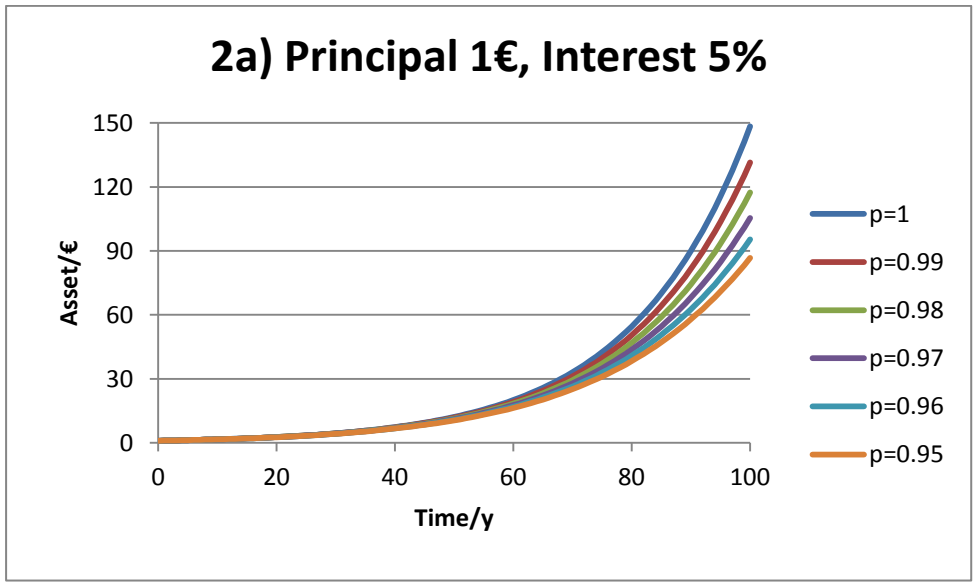

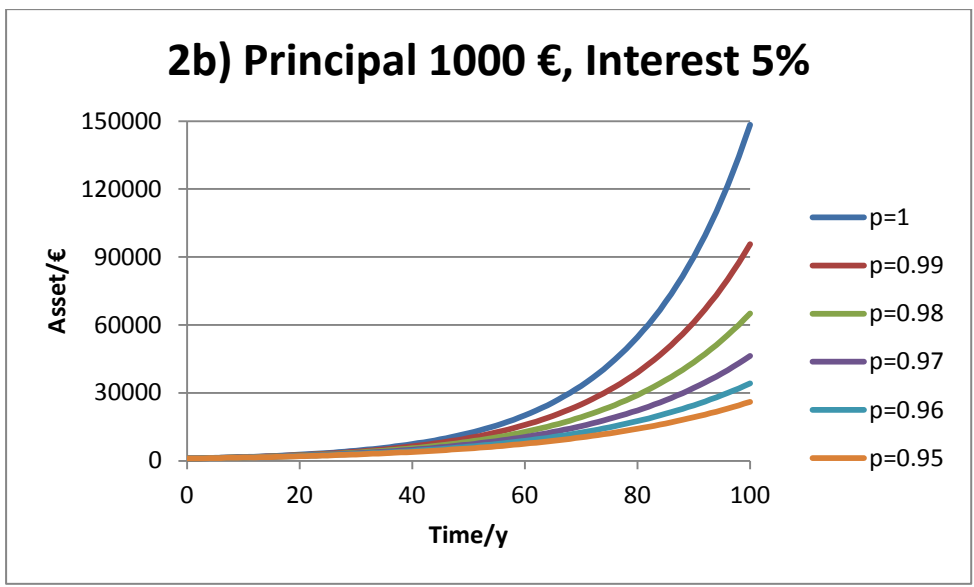

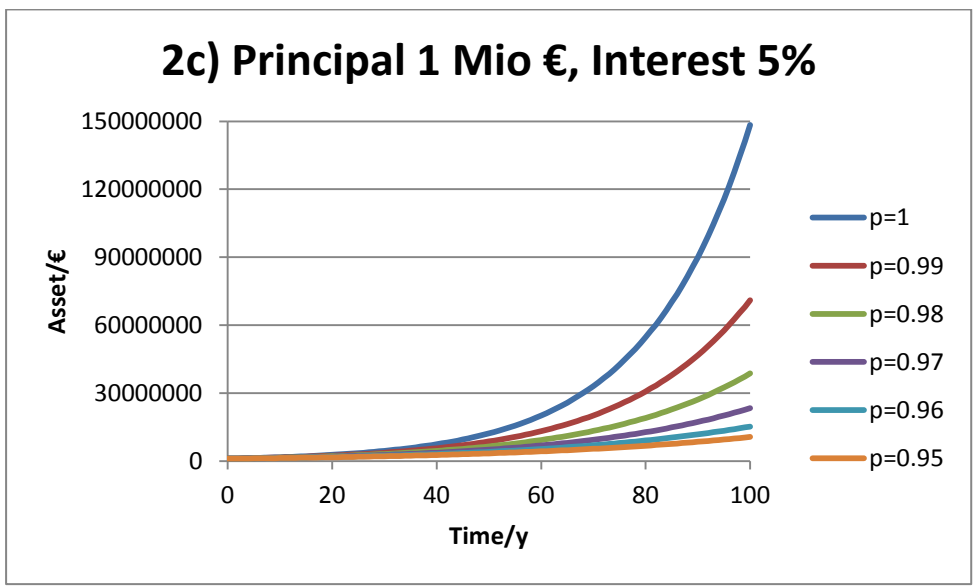

**Figure 2:** Asset growth in the subexponential case for a principal of (a) 1 €, (b) 1000 €, (c) 1 Mio €, (d) 1 Mrd €, a fixed annual interest rate of 5%, and continuous compounding. Monetary growth orders are shown in the legends.

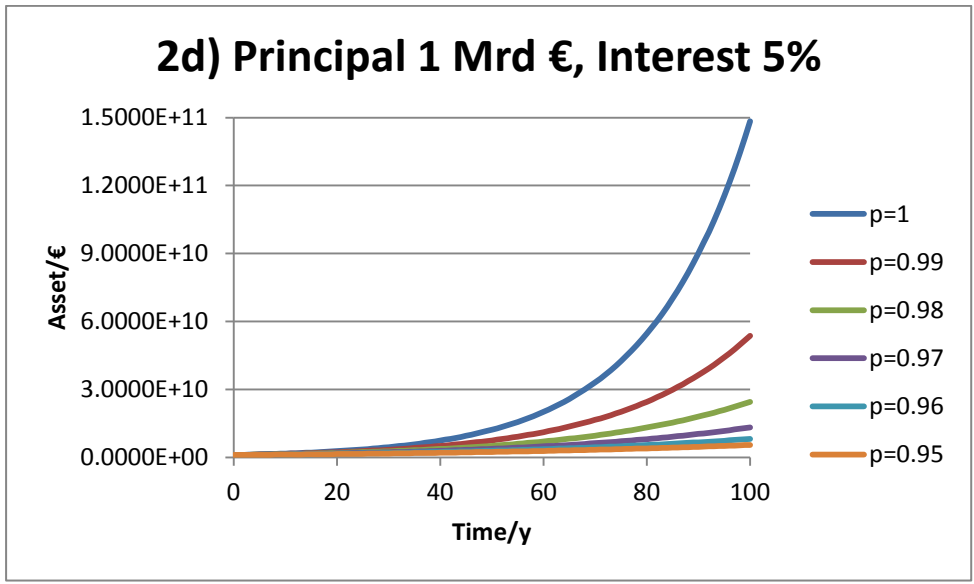

**Figure 2** continued

Supraexponential compounding (Fig. 1a-d) behaves dramatically different. If the monetary growth order p is increased by just 1%, the time needed to reach a factor of 148 shortens for about 2.5 years at a miniasset of 1 €, about 8 years at 1000 €, about 12 years at 1 Mio €, and almost 15 years at 1 Mrd €. Note that this shortening does not correlate with the monetary growth order in a linear fashion. In any case supraexponential growth means that large assets grow faster than smaller ones, both absolutely, and relatively. At a monetary growth order of 1.05 (5% above exponential) and an interest rate of 5%, a milliardaire (US: billionaire) would need less than 30 years to gain an asset which can be compared to the whole European Financial Stabilisation Mechanism of 2011. Clearly, supraexponential growth will not be able to solve the current financial problems – most likely it will never be useful.

Subexponential growth (Fig. 2a-d) however is expected to be. At 5% interest rate the effect of a 1% decrease of the monetary growth order (p = 0.99) reduces our factor of 148 (obtained for exponential growth after 100 years) to a factor of 132 at 1 €, 96 at 1000 €, 71 at 1 Mio €, and 54 at 1 Mrd €. A 5% decrease (p = 0.95) ends at factors 87, 26, 11.6 and 5.5, respectively. Subexponential growth thus achieves that small assets grow faster than larger one on the relative scale, while never outperforming large assets absolutely. As a consequence, large assets will stay large in a subexponential phase – just growing slower here. In other words: The rich will stay rich but the poor will not suffer a further increase of poverty.

Fig. 3a-d gives clues how the asset doubling times depend on interest rate i, monetary growth order p, and the principal $c_0$. Calculations are based on equation (2), by setting $c_t = 2\, c_0$ and resolving for the time:

$$(7) \qquad t_2 = \frac{c_0^{1-p}}{i} \frac{2^{1-p}-1}{1-p}$$

As expected for the case of exponential growth a plot of the doubling time against the interest rate reveals a hyperbola with $t_2$ = 70 years at 1% interest rate and 7 years at 10%. It becomes evident that while small assets increase almost exponentially even at p = 0.95, large assets become more and more sensitive to the effect of the monetary growth order. To achieve a doubling time of 10 years, an asset of 1 Mrd € has to come along with an interest rate of 7% for the exponential case and 8.4%, 10.6%, 13%, 16%, and 20% for monetary growth orders of p = 0.99, 0.98, 0.97, 0.96, and 0.95, respectively. The subexponential compounding mechanism is thus an effective means to stop large assets and debts creating problems of the kind we see today.

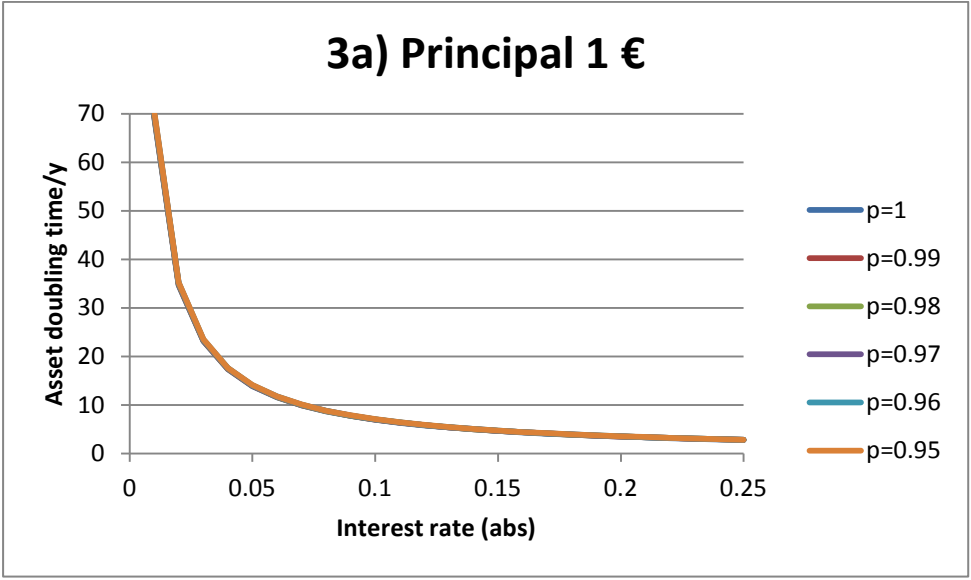

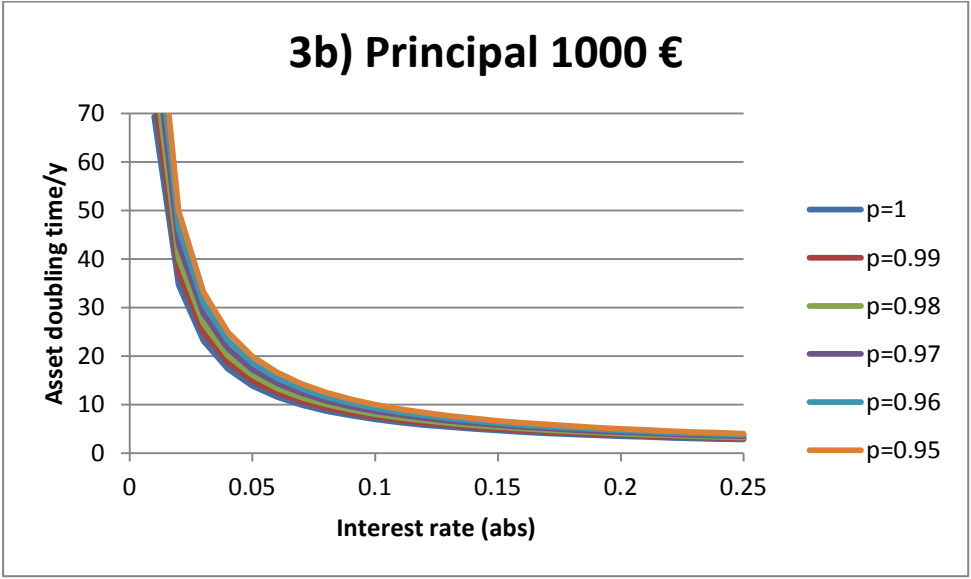

**Figure 3:** Asset doubling time as a function of interest rate for a principal of (a) 1 €, (b) 1000 €, (c) 1 Mio €, (d) 1 Mrd € and the monetary growth orders as shown in the legends.

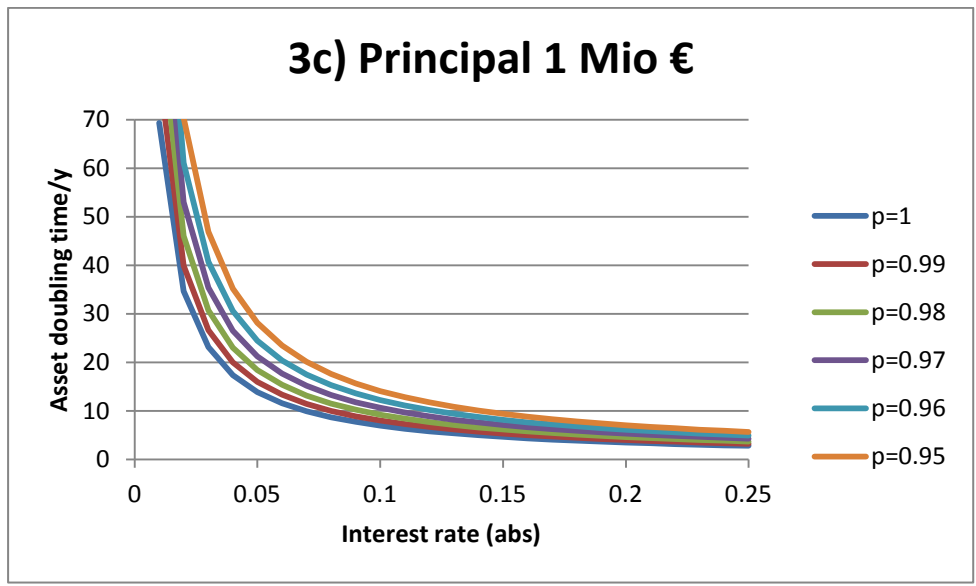

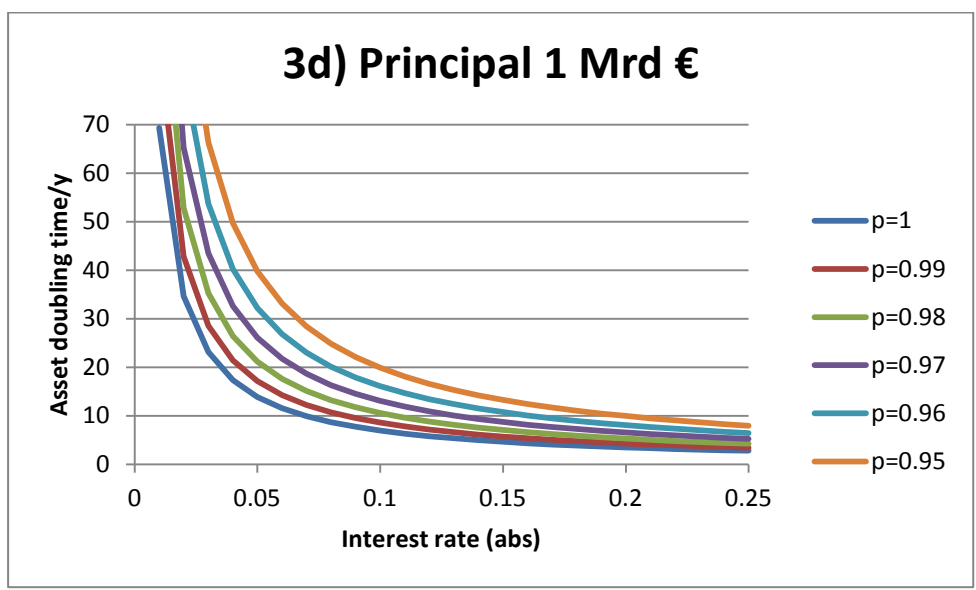

**Figure 3** continued

**Open problems**

Our proposal that a central bank should control the monetary growth order shifts much political responsibility to this bank. Which macroeconomic data are to be employed to set a growth order that improves the overall performance? How do well established macroeconomic equations such as the Fisher equations change upon moving from a fixed exponential mode of compounding to a variable one? How does the monetary growth order affect the Pareto distribution, the Lorenz curve or the Gini index? How do the latter statistical data change with time for an exponential versus non-exponential modes of monetary growth? Should a change in the monetary growth order only affect new credits or should all assets and debts, including the existing ones, continuously feel the actual growth order as a feedback? How to change the mechanics of saldoing [10+]? Is it necessary to introduce, for example, a European tax number in order to assign asset components spread over many bank accounts? Should the correction by the monetary growth order become an issue of merchant banking (as proposed) or taxation (as an alternative)? While the monetary growth order will have an influence in the classical realm of banking based on the creation of credits, it needs further research to understand how the financial markets might respond on its introduction. Could it make sense to see the monetary growth order and a financial transaction tax [10++, 10+++] as complementing instruments aiming to cure a debt crisis? While there are definitely more questions than answers, we feel nevertheless encouraged to propose the monetary growth order, as the well established exponential case of compounding is included in equations (1) and (2) (as a limit case) and thus a return to the status quo is possible at all times. We believe that our proposal may contribute to solve a number of problems which seem to be inherent to a monetary system based on compound interest. The main issue of the current system with a fixed exponential mode of monetary growth is that it automatically leads problems relating to distributive justice and that these problems grow with time finally resulting in a gross financial destabilization and an economic crisis [11, 12, 12+]. Unlike other proposals [13] the introduction of a monetary growth order does not mean a radical deviation from the current practice of finance based on the division of labour between a central bank and connected merchant banks. It definitely needs expert knowledge on the implementation of the monetary growth order, knowledge that goes beyond the general insight chemistry and biology can provide on self-maintaining growth and its evolutionary consequences.

**Outlook**

While students of natural sciences are well advised to avoid too much generalization of a particular insight it occasionally happens that knowledge derived in one field embeds a principle that calls for application in another field. Even the work context of a scientific study may cause a scientist to look over the rim of his field as it happened with Sir Frederick Soddy in the early decades of the last century. Soddy, who won the 1921 Nobel Prize in Chemistry for his work on the alpha-decay of radioactive elements and the discovery of radioactive isotopes switched to economics afterwards. In his books *Wealth, virtual wealth and debt* [14] and *The role of money* [15] which predated J.M. Keynes famous "General theory of employment, money and interest" [16] Soddy addressed the process of creation of giral money, assets and debts booked on banking accounts, and arrived at the conclusion that the exponential nature of compounding and the accumulation of debts create imbalances that could only be resolved by war in these days. Soddy also foresaw that the production factors of classical liberalism (Smith's land, labour and capital) need to be redefined to account for modern economic evolution. In Soddy's view (scientific and technical) discovery, the energy of our sun (either as flux or in its stored form, viz. coal, gas, oil) and human diligence replace the classical production factors.

To quote Wikipedia: "In four books written from 1921 to 1934, Soddy carried on a 'quixotic campaign for a radical restructuring of global monetary relationships', offering a perspective on economics rooted in physics – the laws of thermodynamics, in particular – and was 'roundly dismissed as a crank'. While most of his proposals – 'to abandon the gold standard, let international exchange rates float, use federal surpluses and deficits as macroeconomic policy tools that could counter cyclical trends, and establish bureaus of economic statistics (including a consumer price index) in order to facilitate this effort' – are now conventional practice, his critique of fractional-reserve banking still 'remains outside the bounds of conventional wisdom'."

We believe that the latter critique would be different if a monetary growth order had been existent in Soddy's days.

**Acknowledgements:**

We thank J Michael McBride, Yale University, Imre Kondor, Eötvös University, Stuart Kauffman, University of Vermont, Wim Hordijk, University of Oxford, Daniel R Brooks, University of Toronto, for critical reading and helpful discussion, Dieter Schinzer, COST office, Brussels, for advice in dissemination, and Tobias Plöger, Ruhr-University Bochum, for technical help.